\pdfoutput=1
\documentclass[aps,prl,groupedaddress,superscriptaddress,showpacs,reprint,nofootinbib]{revtex4-1}
\usepackage{amsmath}
\usepackage{hyperref}
\usepackage{graphicx}
\usepackage{stackengine}
\usepackage{color}
\usepackage{subfigure}

\usepackage{helvet}
\usepackage{times}
\usepackage[inline]{enumitem}

\begin{document}

	\title{Solving Graph Problems Using Gaussian Boson Sampling}
	
	
	
	\author{Yu-Hao Deng}
	\thanks{These authors contributed equally to this work.}
	\author{Si-Qiu Gong}
	\thanks{These authors contributed equally to this work.}
	\author{Yi-Chao Gu}
	\thanks{These authors contributed equally to this work.}
	\author{Zhi-Jiong Zhang}
	\author{Hua-Liang Liu}
	\author{Hao Su}
	\author{Hao-Yang Tang}
	\author{Jia-Min Xu}
	\author{Meng-Hao Jia}
	\author{Ming-Cheng Chen}
	\author{Han-Sen Zhong}
	\author{Hui Wang}
	\author{Jiarong Yan}
	\author{Yi Hu}
	\affiliation{Hefei National Laboratory for Physical Sciences at Microscale and School of Physical Sciences, University of Science and Technology of China, Hefei, Anhui, 230026, China}
	\affiliation{CAS Centre for Excellence and Synergetic Innovation Centre in Quantum Information and Quantum Physics, University of Science and Technology of China, Shanghai, 201315, China}
	\affiliation{Hefei National Laboratory, University of Science and Technology of China, Hefei 230088, China}

	\author{Jia Huang}
	\author{Wei-Jun Zhang}
	\author{Hao Li}
	
	\affiliation{State Key Laboratory of Functional Materials for Informatics, Shanghai Institute of Micro system and Information Technology (SIMIT), Chinese Academy of Sciences, 865 Changning Road, Shanghai, 200050, China}
	
	\author{Xiao Jiang}

	\affiliation{Hefei National Laboratory for Physical Sciences at Microscale and School of Physical Sciences, University of Science and Technology of China, Hefei, Anhui, 230026, China}
	\affiliation{CAS Centre for Excellence and Synergetic Innovation Centre in Quantum Information and Quantum Physics, University of Science and Technology of China, Shanghai, 201315, China}
	\affiliation{Hefei National Laboratory, University of Science and Technology of China, Hefei 230088, China}

	\author{Lixing You}
	\author{Zhen Wang}
	
	\affiliation{State Key Laboratory of Functional Materials for Informatics, Shanghai Institute of Micro system and Information Technology (SIMIT), Chinese Academy of Sciences, 865 Changning Road, Shanghai, 200050, China}
	
	\author{Li Li}
	\author{Nai-Le Liu}
	
	\affiliation{Hefei National Laboratory for Physical Sciences at Microscale and School of Physical Sciences, University of Science and Technology of China, Hefei, Anhui, 230026, China}
	\affiliation{CAS Centre for Excellence and Synergetic Innovation Centre in Quantum Information and Quantum Physics, University of Science and Technology of China, Shanghai, 201315, China}
	\affiliation{Hefei National Laboratory, University of Science and Technology of China, Hefei 230088, China}

	\author{Chao-Yang Lu}
	
	\affiliation{Hefei National Laboratory for Physical Sciences at Microscale and School of Physical Sciences, University of Science and Technology of China, Hefei, Anhui, 230026, China}
	\affiliation{CAS Centre for Excellence and Synergetic Innovation Centre in Quantum Information and Quantum Physics, University of Science and Technology of China, Shanghai, 201315, China}
	\affiliation{Hefei National Laboratory, University of Science and Technology of China, Hefei 230088, China}
	\affiliation{New Cornerstone Science Laboratory, Shenzhen 518054, China}
	
	\author{Jian-Wei Pan}
    \affiliation{Hefei National Laboratory for Physical Sciences at Microscale and School of Physical Sciences, University of Science and Technology of China, Hefei, Anhui, 230026, China}
	\affiliation{CAS Centre for Excellence and Synergetic Innovation Centre in Quantum Information and Quantum Physics, University of Science and Technology of China, Shanghai, 201315, China}
	\affiliation{Hefei National Laboratory, University of Science and Technology of China, Hefei 230088, China}

	\date{\today}
	
	\begin{abstract}
   Gaussian boson sampling (GBS) is not only a feasible protocol for demonstrating quantum computational advantage, but also mathematically associated with certain graph-related and quantum chemistry problems. In particular, it is proposed that the generated samples from the GBS could be harnessed to enhance the classical stochastic algorithms in searching some graph features. Here, we use \textit{Ji\v{u}zh\={a}ng}, a noisy intermediate-scale quantum computer, to solve graph problems. The samples are generated from a 144-mode fully-connected photonic processor, with photon-click up to 80 in the quantum computational advantage regime. We investigate the open question of whether the GBS enhancement over the classical stochastic algorithms persists---and how it scales---with an increasing system size on noisy quantum devices in the computationally interesting regime. We experimentally observe the presence of GBS enhancement with large photon-click number and a robustness of the enhancement under certain noise. Our work is a step toward testing real-world problems using the existing noisy intermediate-scale quantum computers, and hopes to stimulate the development of more efficient classical and quantum-inspired algorithms.
    \end{abstract}
	
	
	\maketitle

\begin{figure*}[!htp]
    \centering
    \includegraphics[width = 0.85\linewidth]{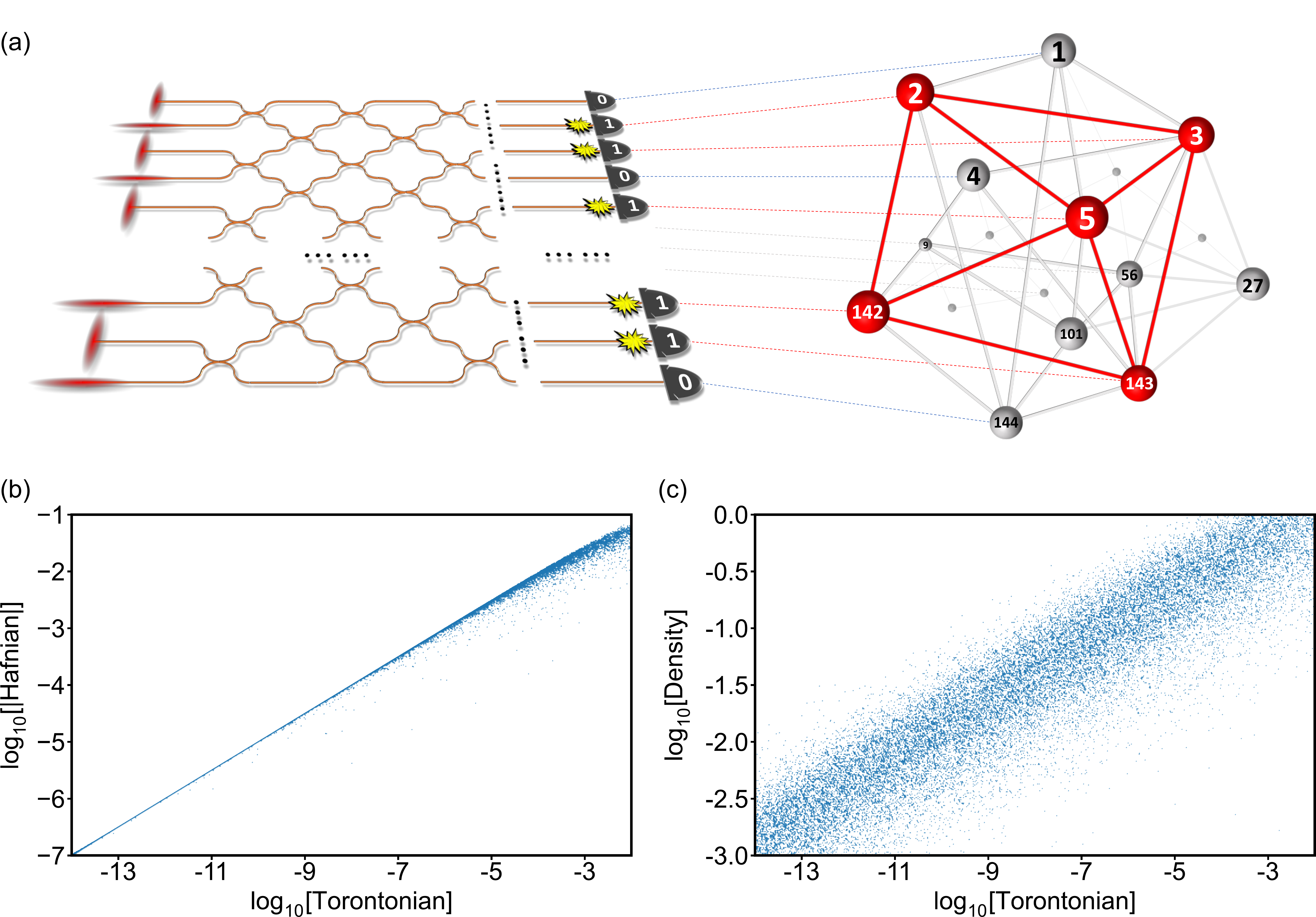}
    \caption{Principle of the experiment. (a) The correspondence between a GBS experiment and a graph. Each output port of the GBS corresponds to a vertex of the graph, and the clicked ports correspond to a subgraph whose vertices and edges are marked in red. (b) The Monte Carlo simulation results showing Hafnian’s dependence on Torontonian for the randomly sampled four-mode complex-valued matrix. (c) The Monte Carlo simulation results showing graph density’s dependence on Torontonian for the randomly sampled four-mode complex-valued matrix.}
    \label{fig:1}
\end{figure*}

Recent experiments have constructed noisy intermediate-scale quantum (NISQ) devices and shown increasingly more convincing evidence for quantum computational advantage \cite{arute2019,zhong2020,zhong2021,wu_strong2021,madsen2022}, a milestone that demonstrates that the quantum devices can solve sampling problems overwhelmingly faster than classical computers. A natural next step is to test whether these NISQ devices can solve problems of practical interest.

Gaussian boson sampling (GBS) \cite{hamilton2017gaussian}, a variant of the original Aaronson-Arkipov boson sampling  \cite{aaronson2011computational}, has attracted considerable attention for its potential applications in graph-related problems, quantum chemistry, and machine learning \cite{huh2015,arrazola_maxhaf_2018,arrazola_dense_2018,bradler_perfectmatching2018,banchi_molecular_2020,jahangiri2021transport,schuld2020similarity}. This is because the GBS has underlying mathematical connection with graph theory. Therefore, theories \cite{arrazola_maxhaf_2018,arrazola_dense_2018,bradler_perfectmatching2018,banchi_molecular_2020} suggested that the generated samples from the GBS might give enhancement over classical stochastic algorithms in solving some graph problems. Moreover, the highly connected topology of the GBS photonic processor can naturally address problems on nonplanar graphs \cite{harrigan2021}.

Proof-of-principle demonstrations of solving graph problems assisted by the GBS have been reported \cite{zhong2019experimental,arrazola2021quantum,paesani2019generation,sempere2022}, however, in regimes where the GBS devices dynamics can be easily simulated on classical computers. An important and open question is whether the GBS could give enhancement on increasingly larger devices in the computationally interesting regime, and how the performance is affected by noise in NISQ devices. Furthermore, previous demonstrations on finding dense subgraphs could only address the  problem with non-negative-valued sampling matrices, for which efficient classical algorithms of estimating the sampling probability exist \cite{Barvinok1999,Rudelson2016} and a quantum-inspired classical algorithm was recently developed \cite{oh2023arxiv}.

Here, we test solving nonplanar graph problems on the NISQ photonic quantum processor, \textit{Ji\v{u}zh\={a}ng}, with 50 single-mode squeezed states input into a 144-mode linear optical network \cite{zhong2020,zhong2021}. We operate \textit{Ji\v{u}zh\={a}ng} in the computationally interesting regime to enhance stochastic algorithms solving two graph problems, namely the Max-Haf problem \cite{arrazola_maxhaf_2018} and the dense $k$-subgraph problem \cite{arrazola_dense_2018}. We  benchmark how the performance scales as a function of the GBS size, and how it is influenced by certain noise \cite{preskillNISQ2018}.

In the GBS, arrays of squeezed vacuum states are sent through a multi-mode interferometer and sample the output scattering events. Due to its Gaussian properties, the output state can be
described by its Husimi covariance matrix $\sigma_{Q}$ \cite{wang2007,weedbrook2012}, for which the sampling matrix is expressed as

\begin{equation}
    \mathcal{A =}\begin{bmatrix}
    0 & I \\
    I & 0 \\
    \end{bmatrix}\left( I - \sigma_{Q}^{- 1} \right).
    \label{eq:1}
\end{equation}
The sampling matrix \(\mathcal{A}\) is in a block matrix form
 \begin{equation}
    \mathcal{A =}\begin{bmatrix}
    A & L \\
    L^{\dagger} & A^{\ast} \\
    \end{bmatrix}.
    \label{eq:2}
 \end{equation}
where $A$ is a symmetric matrix, and \(L = 0\) if the Gaussian state is a pure state. 

An illustration of the correspondence between a graph and a GBS setup is shown in Fig. \ref{fig:1}(a). Any undirected graph can be represented by its adjacency matrix $\Delta$ which is symmetric---i.e. $\Delta_{ij}=\Delta_{ji}$---and the adjacency matrix element $\Delta_{ij}$ corresponds to the weighted value of the edge connecting vertex $i$ to vertex $j$. The adjacency matrix can be encoded into the sampling matrix \(\mathcal{A}\) of a pure state GBS with a proper rescaling factor $c$ (see the Supplemental Material \cite{suppl}):
 \begin{equation}
    \mathcal{A =}\begin{bmatrix}
    c\Delta & 0 \\
    0 & c\Delta^{\ast} \\
    \end{bmatrix}.
    \label{eq:2}
 \end{equation}
and by Takagi-Autonne decomposition \cite{horn2012matrix} the corresponding GBS setup can be constructed. Each mode of the output light field maps to a column and row of the adjacency matrix, and each GBS sample corresponds to a submatrix of the sampling matrix
\(\mathcal{A}\) by taking the elements of the corresponding rows and columns. Once the relationship between the GBS device and the adjacency matrix of the graph under study is established, the GBS samples, whose probability is positively correlated to the
mathematical quantity called \textit{Torontonian} \cite{quesada2018Tor} (for threshold detection) or \textit{Hafnian} \cite{kruse2019detailed} (for photon-number- resolving detection) of the corresponding submatrix, are harnessed to enhance solving the graph problems of interest.

\begin{figure*}[!htp]
    \centering
    \includegraphics[width = 0.59\textwidth]{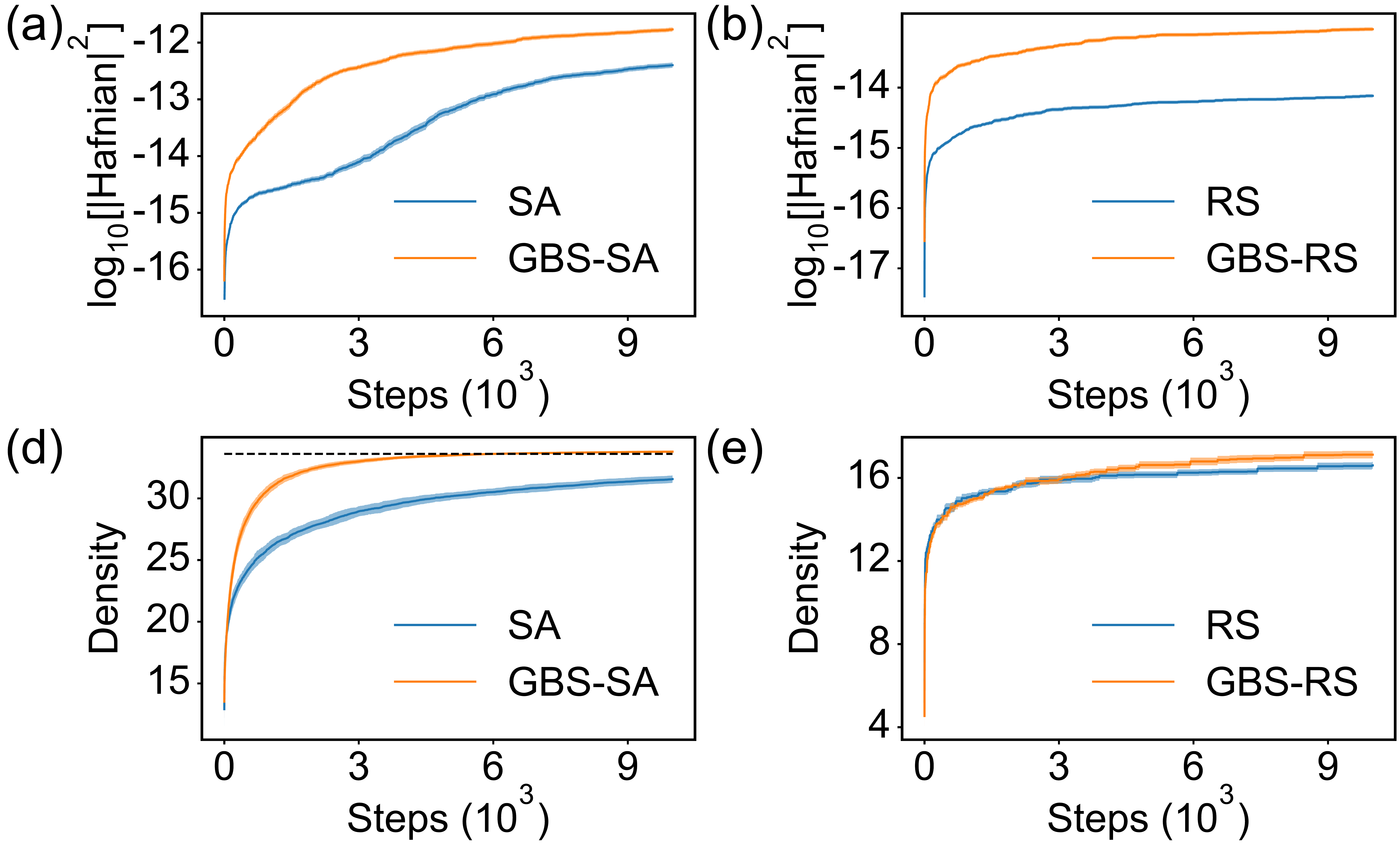}
    \includegraphics[width = 0.4\linewidth]{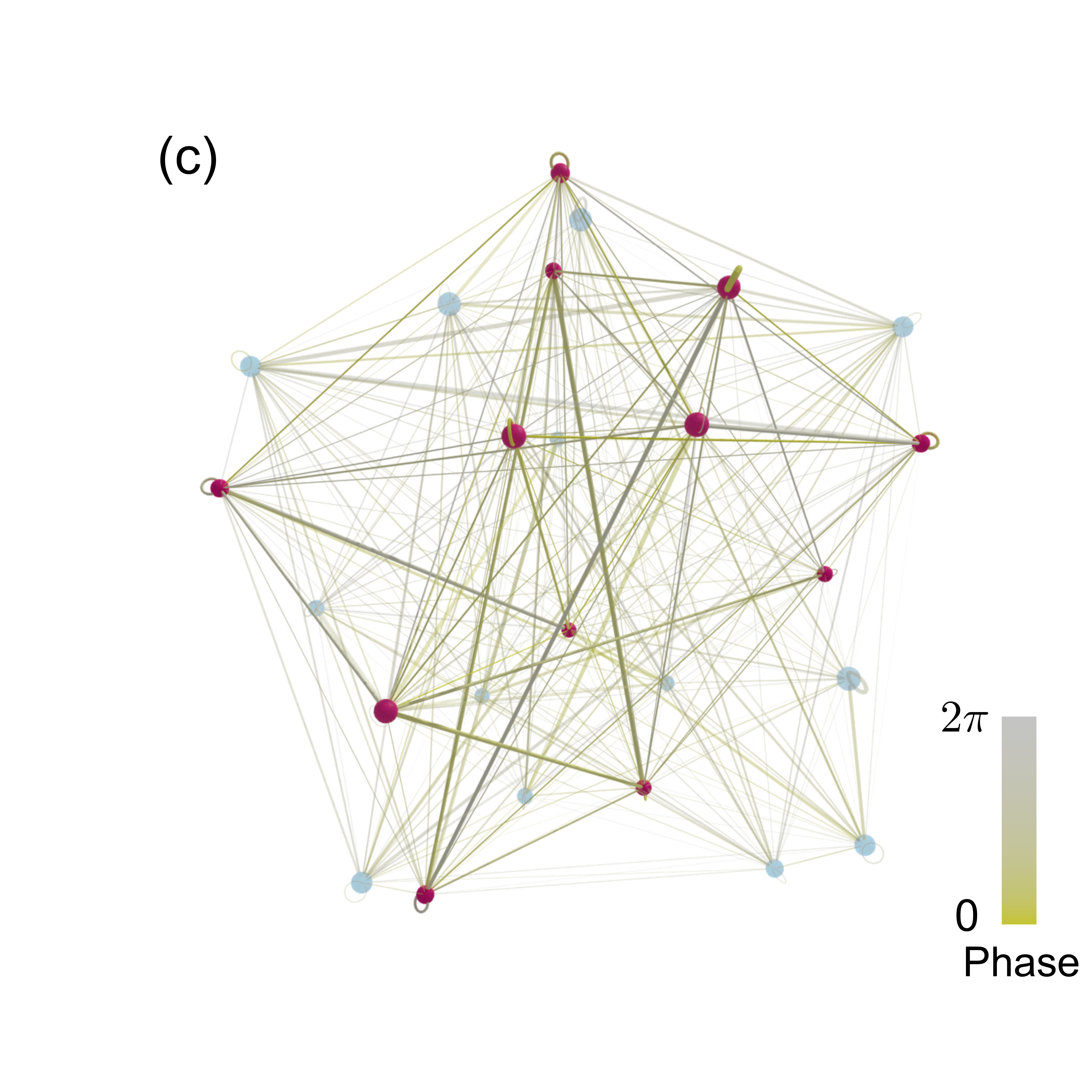}
    \caption{The GBS enhancement on Max-Haf and dense $k$-subgraph problem solving.
    (a,b) The GBS enhancement on finding the 12-vertex subgraph with the largest Hafnian out of a 144-vertex graph for SA (a) and RS (b) algorithm. The experiment obtains a mean photon-click number of 21. A set of 12-photon-click samples are post-selected for this study.
    The X-axis corresponds to the number of steps used in the optimization process, while the Y-axis indicates the corresponding largest Hafnian found. 
    Each curve is the mean largest Hafnian out of 100 trials, and the shaded area indicates standard error.
    (c) The graph corresponding to the experiment, and the subgraph of maximum Hafnian found by the GBS-enhanced SA algorithm (vertices marked in red).  119 unchosen vertices are omitted for the display.
    (d,e) The GBS enhancement on finding the 80-vertex dense subgraph out of a 144-vertex graph for SA (d) and RS (e) algorithm.
    The experiment obtains a mean photon-click number of 61. A set of 80-photon-click samples are post-selected for this study.
    The X-axis corresponds to the number of steps used in the optimization process, while the Y-axis indicates the corresponding largest density found. The curve is the maximum (mean) largest density out of 120 (20) trials for SA (RS), and the shaded area indicates standard deviation (error). In (d) The horizontal dashed line shows the density found by the deterministic greedy algorithm, and it is surpassed by the GBS-enhanced SA algorithm. 
}
    \label{fig:2}
\end{figure*}

We study the GBS enhancement on solving the
Max-Haf problem and dense $k$-subgraph problem. 
The Max-Haf problem is, for a complex-valued matrix $B$ of any dimension, to find a submatrix $B_S$ of fixed even dimension $k=2m$, with the largest Hafnian in square of absolute value. Hafnian was originally introduced in interacting quantum field theory
and plays a variety of roles in physics and chemistry \cite{caianiello1953quantum,fisher1966dimer,heilmann1972dimer,moore2011nature,huh2015,john1998kekule,john1990calculating,hosoya1971topological}. When the matrix is
an adjacency matrix composed of 0s and 1s, Hafnian can be interpreted as
the number of perfect matching of the graph \cite{bradler_perfectmatching2018}. The Max-Haf problem is known to belong to the NP-hard complexity class \cite{arrazola_maxhaf_2018}.

The dense $k$-subgraph problem is, for an $n$-vertex graph $G$ with adjacency matrix $\Delta$, to find its subgraph of $k<n$ vertices $G_S$ with the largest density
  \begin{equation}
     W\left( G_{S} \right) = \left| \sum_{i,j=1}^{k}{(\Delta_S)}_{i,j} \right|,
  \end{equation}
 where $\Delta_S$ is the adjacency matrix of $G_S$.
 
The dense $k$-subgraph problem is of fundamental interest in both
mathematics \cite{feige2001dense} and applied fields like data mining, bioinformatics,
finance and network analysis \cite{kumar1999trawling,angel2012dense,beutel2013copycatch,chen2010dense,fratkin2006motifcut,saha2010dense,arora2011computational,leskovec2008statistical}. Although there are deterministic algorithms for finding subgraph of large density, they can be
fooled and thus stochastic algorithms are important in certain scenarios \cite{arrazola_dense_2018}.

The principle of the GBS enhancement on solving
the two problems by stochastic algorithms can be understood from the
concept of proportional sampling. Since the GBS samples are more likely to have
a larger Hafnian in modulus (hereinafter we use Hafnian to refer to Hafnian in modulus), it also holds that subgraphs
corresponding to the GBS samples are more likely to have larger Hafnian.
Therefore, one can use the GBS samples to boost the effectiveness of stochastic algorithms in
solving the Max-Haf problem by augmenting its success probability. 

Furthermore, it is proved that for a graph
of 0s and 1s, its density is positively correlated to Hafnian, and dense $k$-subgraph problem solving can also be expected to gain enhancement from GBS \cite{arrazola_dense_2018}. From
another point of view, by working in a quantum-classical hybrid scheme, the
GBS serves as an oracle to significantly
narrow down the combinatorial search space of the
stochastic algorithm since the subgraphs of small Hafnian or density are unlikely to be sampled.

These two graph problems differentiate from each other by means of
their target function's computational complexity. Hafnian is hard-to-compute while density can be evaluated efficiently. Investigation on the two graph problems of distinct properties provides us with insights into the dependence of GBS enhancement on the computational complexity of the graph feature itself.

While the above discussion holds for the ideal GBS, in experiments we need to consider three realistic derivations.
\begin{enumerate*}[label=(\roman*)]
\item The sampling matrix retrieved from the experiment is not always ideally non-negative as in the original proposal. Imperfection in circuits can introduce negative or imaginary terms into the sampling matrix.
\item Experimental noise like photon loss causes mixed state sampling in GBS, which can result in biased diagonal block matrix $A$, and nonzero off-diagonal block matrix \(L \neq 0\) \cite{qi2020regimes}.
\item Threshold detectors are usually used instead of photon-number-resolving detectors \cite{hadfield2009detector}.
\end{enumerate*}

To check whether the proportional sampling mechanism holds for the GBS with threshold detectors on complex-valued sampling matrices, we perform Monte
Carlo simulation to reveal the numerical correlation between Torontonian
and Hafnian or density for randomly generated complex sampling matrices (see the Supplemental Material \cite{suppl}). Shown in Fig.
\ref{fig:1}(b) and (c), the positive correlation between Torontonian and Hafnian or
density validates the underlying principle of proportional sampling, and portends the occurrence of GBS enhancement. 

\begin{figure*}[!htp]
    \centering
    \includegraphics[width = 0.76\textwidth]{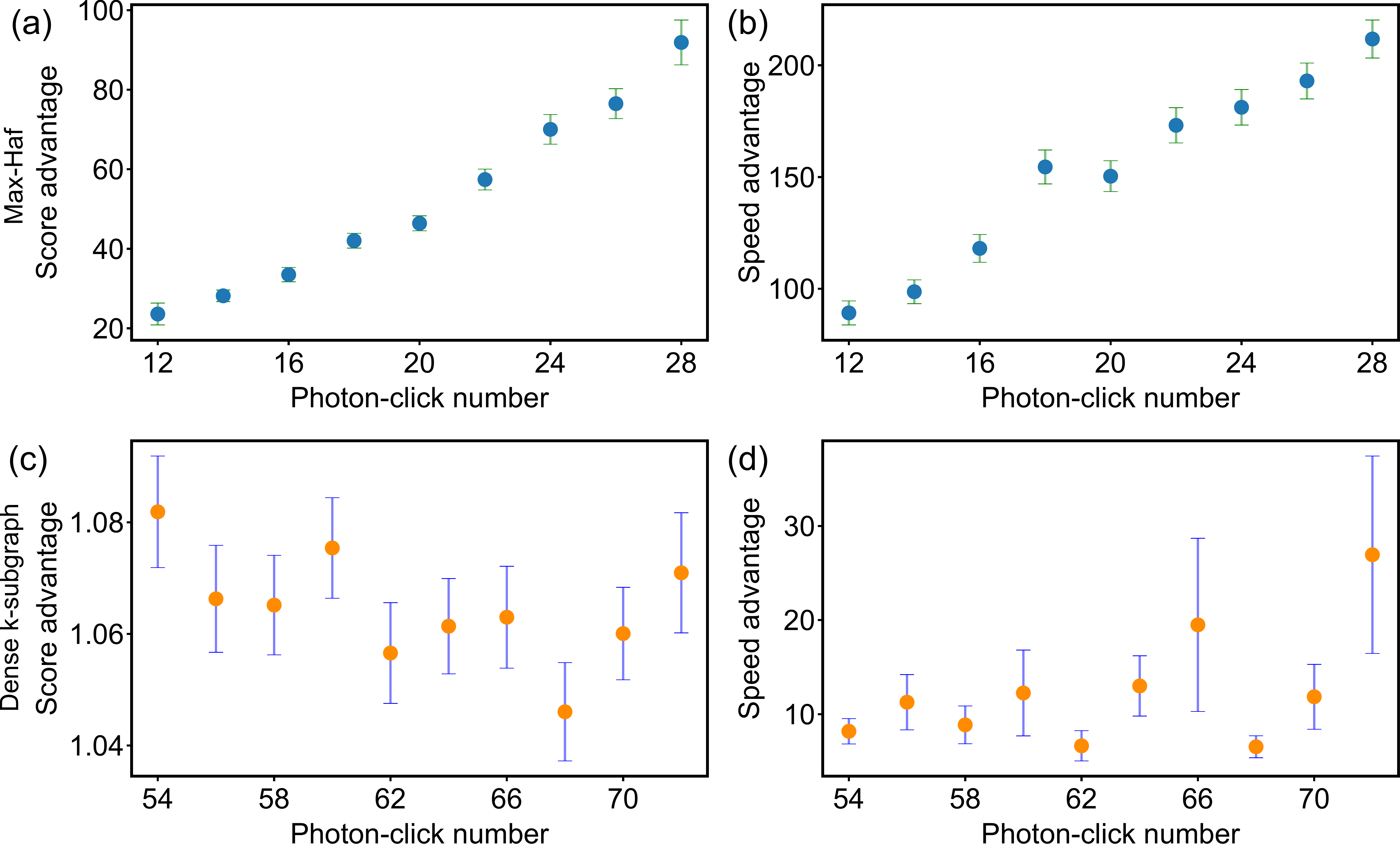}
    \caption{The scaling of the GBS enhancement benchmarked by the score advantage and speed advantage as a function of photon-click number. The problem is defined on a randomly generated complex-valued full graph of 144 vertices. (a),(b) The GBS enhancement on the Max-Haf problem with various photon-click numbers.
    The mean photon-click number of the experiment is 21.
    The score advantage as a function of photon-click number is shown in (a), which is defined as the ratio of maximum Hafnian in square of modulus searched at 1000 steps by the GBS-enhanced RS algorithm to that searched by the RS algorithm. The speed advantage, which is defined as the ratio of the number of steps reaching the target value by the RS algorithm to that by the GBS-enhanced RS algorithm, is shown in (b) as a function of photon-click number. The target value of each trial is set as that reached by the RS algorithm at 1000 steps. A clear rising trend with increased photon-click number can be observed for both the score advantage and the speed advantage.
    (c),(d) The GBS enhancement on the dense $k$-subgraph problem for various photon-click number. 
    The mean photon-click number of the experiment is 61.
    The score advantage is displayed in (c), which is defined as the ratio of the density optimized at 10000 steps by the GBS-enhanced RS algorithm to that by the RS algorithm, as a function of photon-click number. The speed advantage which is defined as the ratio of the number of steps reaching the same density by the RS algorithm to that by the GBS-enhanced RS algorithm, is showns in (d) for various photon-click numbers. For each trial, the target value is set as that reached by the the RS algorithms at 10000 steps. No significant increasing trending with photon-click number is observed for this problem. Error bars indicate standard error.
    }
    \label{fig:3}
\end{figure*}

We proceed to test the GBS enhancement on solving
Max-Haf problem and dense $k$-subgraph problem. Two stochastic algorithms---namely, the random search (RS)
and simulated annealing (SA) \cite{arrazola_dense_2018,arrazola_maxhaf_2018}---are studied. RS represents the naive way of solving the combinatorial problem by uniformly sampling from the whole solution space, which is free from being trapped by local optimum but is costly and inefficient. SA combines mechanisms from both random exploration that prevents it from being stuck in local minima and hill climbing that enables it to approach good solutions fast, but proper choice of parameters is crucial for guaranteeing the algorithm's performance. Together, the two algorithms of distinct working subroutines help benchmark the enhancement of GBS on graph applications more comprehensively.

The experiment is performed on a randomly generated and fully connected 144-mode optical interferometer, and a subset of samples with coincident photon-click number up
to 80 are used for the study. Figure \ref{fig:2}(a) and (b) show the maximum Hafnian of a 12-vertex
subgraph on a 144-vertex full graph found for the two algorithms and their GBS-enhanced variants as a function of searching steps. For both the RS and SA algorithm, it is evident the GBS-enhanced variants improve the effectiveness of the algorithms by finding larger Hafnian within the same steps. An illustration of the full graph corresponding to the experiment, together with the subgraph searched by the GBS-enhanced SA algorithm in highlight, is shown in Fig. \ref{fig:2}(c). 

Similarly, Figure \ref{fig:2}(d) and (e) plot the largest
density found at various steps for the four algorithms. The GBS samples are 80 photon-click events from the 144-mode quantum device, which are in the quantum advantage regime. On average each sample would take \textit{Frontier}, the current fastest supercomputer in the world \cite{top500}, more than 700 seconds to generate using exact methods, as estimated with the state-of-the-art classical sampling algorithm \cite{bulmer2022boundary}, and we used 221891 samples in total for the study which amount to $\sim5$ years on \textit{Frontier}. It is observed that both RS and SA algorithms gain enhancement from the GBS samples in searching for subgraphs of higher density at the given step. Specially, it is noted that the density found by the deterministic Greedy algorithm \cite{charikar2000greedy}, which is marked as the horizontal dashed line, can be outperformed by the GBS-enhanced SA algorithm, confirming the advantage of stochastic algorithms.

\begin{figure*}[!htp]
    \centering
    \includegraphics[width = 1\linewidth]{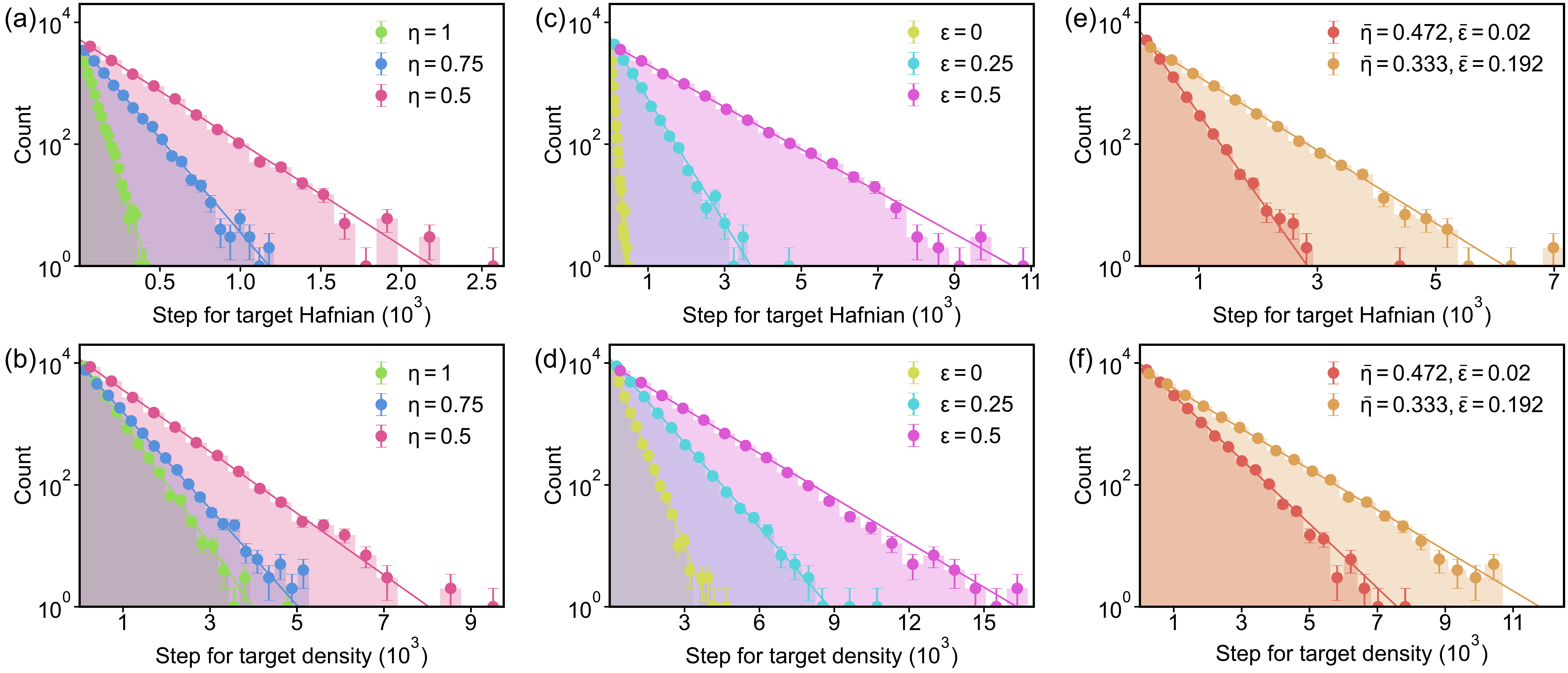}
    \caption{Histogram of steps for GBS-enhanced RS algorithms to achieve the target value as a function of various noise levels. For each subplot, a number of trials are repeated, and the $y$ axis corresponds to the counts of the occurrence of the number of steps to surpass the target value. The target is set as the mean value reached by the classical RS algorithm at 10000 steps. 
    (a),(b) Theoretical simulation of the effect of photon loss noise on the GBS enhancement on the Max-Haf problem (a) and the dense $k$-subgraph problem (b). The plots show the GBS enhancement with samples from a theoretically simulated sampler with no photon loss $\eta=1$, and theoretically simulated sampler with system efficiency $\eta=0.75$ and $\eta=0.5$. It is observed that the ideal sampler without photon loss needs fewer steps for the task than the lossy sampler.
    (c),(d) Simulating thermal noise's influence on GBS-enhancement for (c) the Max-Haf problem and (d) the dense $k$-subgraph problem. The plots exhibit the GBS enhancement with samples from a theoretically simulated sampler of zero thermal noise $\epsilon=0$, and theoretically simulated samplers of thermal noise level $\epsilon=0.25$ and $\epsilon=0.5$. The results indicate that the simulated sampler with no thermal noise needs fewer steps to achieve the target value.
    (e),(f) The GBS-enhancement with samples from a low noise level experiment (red) and a controlled high noise level experiment (brown) on (e) the Max-Haf problem and (f) the dense $k$-subgraph problem. Fewer steps are needed by the low noise level experiment than the high noise level experiment. The low noise level experiment corresponds to an averaged system efficiency $\bar{\eta}=0.472$ and averaged thermal noise level $\bar{\epsilon}=0.02$, whereas the high noise level corresponds to an averaged system efficiency $\bar{\eta}=0.333$ and an averaged thermal noise level $\bar{\epsilon}=0.192$.
    For all plots the problem is defined on a 144-vertex full graph with a photon click of 10. The low- (high-) noise level experiment has a mean photon number of 7 (6), and samples of photon click 10 are used for the investigation. Error bars indicate statistical fluctuation.
}
    \label{fig:4}
\end{figure*}

Having established the GBS enhancement, we continue to investigate how this enhancement scales on our device. We benchmark the GBS enhancement by defining the score advantage and speed advantage. The former is that, for a given step, the maximal score (in terms of the Hafnian or the density) obtained by the GBS-enhanced algorithms divided by that by the classical algorithms only. The latter is that, to reach a target score, the ratio of the needed searching steps between the GBS-enhanced and classical algorithms \cite{suppl}.
We use the parameter-free RS algorithm to probe the scaling properties.

Figure \ref{fig:3}(a) shows the scaling of the score advantage of the GBS enhancement for a fixed $10^3$ steps. Remarkably, the score advantage rises up steadily, from $\sim24$ at a photon click of 12 to $\sim92$ at a photon click of 28. The speed advantage is plotted in Fig. \ref{fig:3}(b) also as a function of photon-click number. The speed advantage starts at $\sim89$ at a photon click of 12 and becomes increasingly larger as the size increases, reaching $\sim212$ at 28 photons. Here, due to the computational overhead, we use the \textit{Sunway TaihuLight} supercomputer to evaluate the Hafnian. Overall, the results of Fig. \ref{fig:3}(a,b) provide strong evidence that the GBS enhancement as benchmarked by the score advantage and speed advantage increases with the photon-click number in solving the Max-Haf problem by RS algorithm on \textit{Ji\v{u}zh\={a}ng}.

Figure \ref{fig:3}(c) plots the score advantage for the dense $k$-subgraph problem at increasing photon-click number. While all the data points show a positive advantage ($>$1), there is no obvious increasing trend at larger size. A similar behavior is observed in the speed advantage, as shown in Fig. \ref{fig:3}(d). In the Supplemental Material \cite{suppl} we show numerically simulated results for an ideal GBS sampler on the dense $k$-subgraph problem of both randomly generated non-negative-valued graph and complex-valued graph, which exhibit trending of the score advantage and speed advantage similar to that reported in our experiment.

Noise is a major problem for the NISQ device. In the GBS, photon loss \cite{sempere2022} (which can be caused by limited efficiency of the optical elements and detection) and thermal noise \cite{qi2020regimes} (which can be caused by spatial mode mismatch of the sources) can turn the pure-state GBS into mixed-state GBS. For the graph problem solving, these noises can make the sampling matrix deviate from the ideal \cite{kruse2019detailed}, and could decrease the positive correlation between Hafnian or density of the encoded matrix and that of the sampling matrix. To characterize the influence of these noises on GBS enhancement on the graph problem solving, we benchmark them with the RS algorithm that is free from parameter choosing. We compare the steps needed for achieving a target value of the problem between samplers of various noise levels. The probability distribution of the steps follows the Geometric Distribution, which gives the probability that the first occurrence of success requires $k$ independent trials.
\begin{equation}
    P(X=k)=(1-p)^{k-1}p
\end{equation}
where $k=1,2...$ is the number of steps, and $p$ is the probability that a GBS sample can produce a better result than the target. The noise's influence on the GBS enhancement can be simply benchmarked by the parameter $p$, since a larger $p$ would indicate fewer steps are needed which correspond to a stronger GBS enhancement, and vice versa.

To investigate the effect of photon loss, we theoretically simulate the performance with an ideal sampler and that with an overall photon loss of $25\%$ and $50\%$, for the same optimization task. Figure \ref{fig:4} (a) and (b) show histograms of the number of steps for the GBS-enhanced RS algorithms to achieve the target value. There is a significant reduction of the steps at increasing system efficiency $\eta$. The $p$ value of the sampler with a unit efficiency is 0.0196 (0.0024) for the Max-Haf (dense $k$-subgraph) problem, whereas the lossy sampler with efficiency $\eta=0.5, 0.75$ correspond to a $p=0.0039$  ($0.0012$), $p=0.0071$  ($0.0018$), respectively. The results indicate that lower photon loss will lead to a stronger GBS enhancement. A recent theoretical study reported similar findings, including the dependence of the GBS enhancement on the partial photon distinguishability \cite{solomons2023arxiv}.

Figure \ref{fig:4} (c) and (d) show the theoretical simulation results for the thermal noise \cite{denggbsthermal}. Three examples are studied, where the thermal noise is chosen for $\epsilon=0, 0.25, 0.5$. Again, a strong decrease of the required steps is observed for lower thermal noise. The $p$ value of the ideal sampler for the Max-Haf (dense $k$-subgraph) problem is 0.0194 (0.0024), whereas the $p$ value of the sampler with $0.25, 0.5$ thermal noise is 0.0023 (0.0011), and 0.0008 (0.0006), respectively. The results show the importance of eliminating the thermal noise to achieve a higher GBS enhancement.

Having studied these effects theoretically, we now benchmark the noise influence experimentally. The experiment results at a typical noise level, where $\bar{\eta}=0.472$ and $\bar{\epsilon}=0.02$, is compared with a controlled higher noise level, $\bar{\eta}=0.333$ and $\bar{\epsilon}=0.192$. As shown in Fig. \ref{fig:4}(e) and (f), the experimental samples with low noise level demonstrate stronger GBS enhancement for both graph problems, which is in good agreement with the theoretical simulation. The $p$ value for the low noise experimental sampler on the Max-Haf (dense $k$-subgraph) problem is 0.0031 (0.0012), whereas it is 0.0014 (0.0008) for the controlled high noise experimental sampler. Interestingly, samples from the modest-noise-level experiments, though with less enhancement, can still improve RS algorithm.

In this Letter, we have demonstrated the GBS enhancement on stochastic algorithms in solving two graph problems of distinct properties with the 144-mode NISQ device \textit{Ji\v{u}zh\={a}ng} in the quantum computational advantage regime. It is an open question, however, whether the GBS can yield advantage compared to improved classical algorithms and quantum-inspired algorithm. Also, the GBS enhancement can depend on the properties of the input graphs, for which more comprehensive algorithm analysis and discussions for various situations are expected. We hope that our work will stimulate experimental efforts on larger-scale, higher-fidelity and fully programmable GBS, exploration of real-world applications where the computational problems can be mapped onto the GBS, and development of more efficient classical and quantum-inspired algorithms.

\begin{acknowledgments}
This work was supported by the National Natural Science Foundation of China, the National Key R\&D Program of China (2019YFA0308700), the Chinese Academy of Sciences, the Anhui Initiative in Quantum Information Technologies, the Science and Technology Commission of Shanghai Municipality (No. 2019SHZDZX01), Innovation Program for Quantum Science and Technology (No. ZD0202010000), the New Cornerstone Science Foundation.

\end{acknowledgments}

%

\newpage
\newpage

\end{document}